\documentclass[10pt,notitlepage,showpacs,pra,twocolumn]{revtex4-1}
\usepackage{ulem}
\usepackage{amsfonts}
\usepackage{amsmath}
\usepackage{amssymb}
\usepackage{graphicx}
\usepackage{float}
\usepackage{color}
\usepackage[toc,page,header]{appendix}

\begin{document}

\title{Anomalous excitation enhancement with Rydberg-dressed atoms}
\author{Xiaoqian Chai, Lu Zhang, Dandan Ma, Luyao Yan, Huihan Bao, Jing Qian$^{\dagger}$ }
\affiliation{Department of Physics, School of Physics and Material Science, East China
Normal University, Shanghai 200062, People's Republic of China}

\begin{abstract}
We develop the research achievement of recent work [M. G\"{a}rttner, \textit{et.al.}, Phys. Rev. Letts. \textbf{113}, 233002 (2014)], in which an anomalous excitation enhancement is observed in a three-level Rydberg-atom ensemble with many-body coherence. In our novel theoretical analysis, this effect is ascribed to the existence of a quasi-dark state as well as its avoided crossings to nearby Rydberg-dressed states. Moreover, we show that with an appropriate control of the optical detuning to the intermediate state, the enhancement can evoke a direct facilitation to atom-light coupling that even breaks through the conventional $\sqrt{N}$ limit of strong-blockaded ensembles. As a consequence, the intensity of the probe laser for intermediate transition can be reduced considerably, increasing the feasibility of experiments with Rydberg-dressed atoms.
\end{abstract}
\email{jqian1982@gmail.com}
\maketitle
\preprint{}

\section{Introduction}
Dressing atoms to Rydberg states (so-called Rydberg-dressed atoms) promises strong and coherent long-range interactions for up to tens of seconds \cite{Honer10,Gaul16,Helmrich16}, making them ideal research candidates in the fields of quantum simulation \cite{Schaub12,Keating13,Schaub15,Saffman16} and especially for strong-correlated systems \cite{Pupillo10,Zhang17}.
Besides, recent researches suggest that using Rydberg-dressed atoms to explore a variety of novel physics, for example, synthetic quantum magnets \cite{Glaetzle14,Glaetzle15,Bijnen15}, ultracold chemical reactions \cite{Wang14}, quantum entanglement of atoms \cite{Mobius13,Schempp15,Jau16}, and nonclassical state of atomic motion \cite{Buchmann17}.
The dressed atoms can be produced via employing a two-photon excitation with a large detuning to intermediate state or a direct off-resonant excitation via a single strong laser field, which have been experimentally realized with atoms trapped individually \cite{Jau16} as well as trapped in optical lattice \cite{Zeiher16}.

For a two-photon excitation scheme as applied in most current experiments, a large atom-light coupling strength, required for generation of Rydberg-dressed atoms with sufficiently strong interactions, is challenging to achieve \cite{Gaul16}. Recent researches have shown that collective excitation of Rydberg states in a strong-blockaded ensemble acquires a $\sqrt{N}$ enhancement to the atom-light coupling strength \cite{Stanojevic09,Lukin01,Heidemann07,Gaetan09,Urban09,Pritchard10,Dudin12,Barrendo14}, suggesting a promising solution to this problem.
The origin of enhancement can be traced back to a large effective Rabi frequency of collective Dicke state guaranteed by Rydberg blockade effect \cite{Dicke54}. Besides, in order to enhance the Rydberg fraction in the dressed atoms, an optional way is offsetting the interaction-induced level shift by a proper optical detuning, which is also known as anti-blockade effect \cite{Ates07,Amthor10,Su17}. More recently, an anomalous excitation facilitation induced by inhomogeneous broadening on the blue-detuned side was proposed in an attractively-interacting Rydberg ensemble \cite{Letscher17}.

In this work, we investigate anomalous excitation facilitation effect in a resonant two-photon excitation system, implemented by an appropriate control of detuning to intermediate state \cite{Rao14}. Different from the previous work where G\"{a}rttner {\it et.al.} first found this effect and attributed it to the buildup of many-body coherence induced by coherent multi-photon excitation between collective states \cite{Garttner14}, we reveal that the essential origin of the facilitation effect is existing an approximate dark eigenstate, with avoided crossings (ACs) to nearby Rydberg dressed states.
Since the Rydberg dressing here is ensured by the electromagnetically induced transparency (EIT) condition, that the probe laser between the ground and intermediate states is kept weaker than the coupling laser between the intermediate and Rydberg states \cite{Fleischhauer05,Parigi12,Weatherill08,Ates11}, then a robust enhanced Rydberg excitation is observable when the detuning is adjusted near ACs, irrespective of the strength of interatomic interactions.

Moreover,  we find the consequent enhancement of the effective Rabi frequency between ground and single-excitation collective state can far exceed usual $\sqrt{N}$ limit, yielding a big reduction to the intensity of probe laser. Different from the result obtained in Ref. \cite{Garttner14} that the dissipation (denoted by rate $\Gamma$) from intermediate state would destroy the enhanced excitation if $\Omega/\Gamma<1.27$ ($\Omega$ is the Rabi frequency for a single atom), we stress that an appropriate control for the detuning can significantly overcome this limitation, extending the enhancement effect into the regime of strong dissipations. In addition, with more atoms included into collective excitation the enhancement will be further significant, which qualifies Rydberg-dressed atoms as an efficient platform for studying long-range interactions \cite{Singer04} and exotic correlated quantum phases \cite{Henkel10,Li17}.

\section{Dark state and avoided Crossings} 

\begin{figure}
\includegraphics[width=3.4in,height=4.7in]{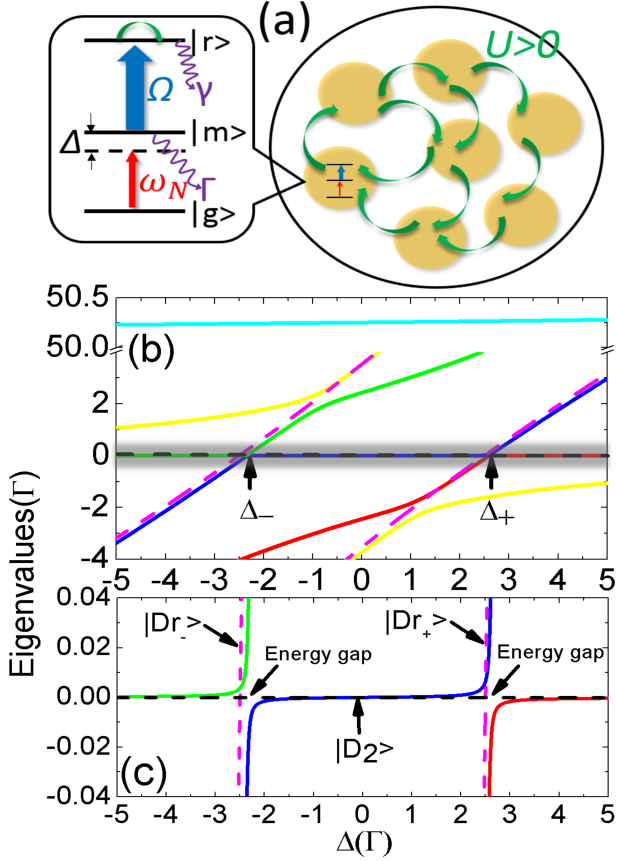}
\caption{(color online). (a1) Schematic view of the atomic ensemble. Each atom has a three-level cascaded configuration. See main texts for a detailed description. (b) The energy of Rydberg-dressed states $\left\vert Dr_{\pm}\right\rangle$ (magenta dashed lines) and dark state $\left\vert D_{2}\right\rangle$ (black dashed line) are shown as functions of detuning $\Delta$. The colored solid lines are for the eigenvalues obtained from numerical diagonalization of Hamiltonian $H_2$. Two ACs are  denoted at $\Delta_{+}$ and $\Delta_{-}$. The parameters are $U=50$, $\Omega=5.0$, $\xi_{2}=0.1$ ($\omega_2=0.5$). $\Gamma$ is the frequency unit. (c) gives a partially enlarged view of (b) (shaded area) where two ACs and energy gaps are clearly distinguishable.}
\label{crossings}
\end{figure}

As illustrated in Fig.\ref{crossings}(a), our ensemble comprises $N$ atoms with three-level cascaded configuration: the ground state $\left\vert g \right\rangle$, the intermediate state $\left\vert m \right\rangle$ and the Rydberg state $\left\vert r \right\rangle$. 
$\left\vert g \right\rangle$ and $\left\vert m\right\rangle$ are coupled by a weak probe laser of Rabi frequency $\omega_{N}$ (the subscript $N$ stands for $N$-atom case) and single-photon detuning $\Delta$, while $\left\vert m \right\rangle$ and $\left\vert r\right\rangle$ are resonantly coupled by a strong laser of Rabi frequency $\Omega$. Incoherent dissipative processes due to spontaneous decay are denoted by the rates $\Gamma$ of $\left\vert m\right\rangle$ and $\gamma$ of $\left\vert r\right\rangle$, respectively, satisfying the condition $\Gamma\gg\gamma$ (Rydberg state $\left\vert r \right\rangle$ is long-lived). 

In the interaction picture the associated Hamiltonian of the single atom $i$ reads ($\hbar=1$)
\begin{equation}
H^{(i)}=\Delta\sigma_{mm}^{(i)}+\frac{\omega_{1}}{2}(\sigma_{gm}^{(i)}+\sigma_{mg}^{(i)})+\frac{\Omega}{2}(\sigma_{mr}^{(i)}+\sigma_{rm}^{(i)}) 
\label{Hamsingle}
\end{equation}
where the atomic operators $\sigma_{\alpha\beta}^{(i)}=\left\vert \alpha_{i}\right\rangle\left\langle \beta_{i}\right\vert$ with $\alpha,\beta\in\{g,m,r\}$. Diagonalizing the single-atom Hamiltonian $H^{(i)}$ leads to a dark eigenstate $\left\vert D_{1}\right\rangle =  \Omega\left\vert g_i\right\rangle - \omega_{1}\left\vert r_i\right\rangle$, and the corresponding Rydberg-state population is 
\begin{equation}
f_{r} = \frac{\omega_{1}^{2}}{\Omega^{2}+\omega_{1}^{2}}
\label{frvalue}
\end{equation}

For a weak probe laser $\omega_{1}<\Omega$, $f_{r}$ is small. In the limit $\omega_{1}\ll\Omega$, a Rydberg-dressed ground state is prepared by weakly dressing the atomic ground state to the Rydberg level \cite{Johnson10,Macri14}. Here note that $f_{r}$ is irrespective of the detuning $\Delta$ as long as the atom is persistently kept on $\left\vert D_{1}\right\rangle$ during the evolution, in other words, increasing $f_{r}$ by adjusting the values of $\Delta$ is of no effective in the case of single atom.

Now we turn to consider an ensemble of $N$ three-level atoms confined in a small volume, concentrating on the effect of interatomic interactions. The $N$-atom Hamiltonian can be written as
\begin{equation}
H_{N} =  \sum_{i=1}^{N}H^{(i)}+U\sum_{i\neq j}^{N}\sigma_{rr}^{(i)}\sigma_{rr}^{(j)}
\label{Ham}
\end{equation}
with $\omega_{1}$ in $H^{(i)}$ replaced by $\omega_{N}$ for the $N$-atom probe laser and $U$ the strength of Rydberg-mediated van der Waals (vdWs) interaction as two atoms occupy the Rydberg states simultaneously \cite{Beguin13,Thaicharoen15}. $U$ is assumed to be positive, see Fig. \ref{crossings}(a).
For two interacting atoms prepared in the ground state [the many-atom case will be discussed in section \rm{V}], it is sufficient to consider the Hamiltonian $H_{2}$ with a set of symmetrical two-atom bases, that is, $\Phi\equiv\{\left\vert gg\right\rangle,\left\vert gm\right\rangle_+,\left\vert gr\right\rangle_+,\left\vert mm\right\rangle,\left\vert mr\right\rangle_+,\left\vert rr\right\rangle\}$ with the definitions $\left\vert
\alpha \beta \right\rangle = \left\vert \alpha \right\rangle \otimes
\left\vert \beta \right\rangle $ and $\left\vert
\alpha \beta \right\rangle_{\pm}=(\left\vert
\alpha \beta \right\rangle\pm\left\vert
\beta\alpha \right\rangle)/\sqrt{2}$.
In the limit $\xi_{2}\equiv\omega_{2}/\Omega\ll 1$ and $U\gg\Omega$, we can diagonalize $H_2$ to the first order of $\xi_2$, which directly gives rise to an approximate two-atom dark eigenstate $\left\vert D_2\right\rangle\approx\left\vert gg\right\rangle$, as well as two Rydberg-dressed eigenstates $\left\vert Dr_{\pm}\right\rangle\approx a_{1\pm}\left\vert mr\right\rangle_{+}+a_{2\pm}\left\vert mm\right\rangle$. Here, the coefficients $a_1$ and $a_2$ are expressed as
\begin{widetext} 
\begin{eqnarray}
a_{1\pm}&=& \frac{\Omega(\mp7\Delta+\sqrt{\Delta^2+2\Omega^2})}{(2(32\Delta^2+\Omega^2)(\Delta^2+2\Omega^2)\mp2\Delta(32\Delta^2+15\Omega^2)\sqrt{\Delta^2+2\Omega^2})^{1/2}}, \label{a1}\\
a_{2\pm} &=& \frac{4\Delta^2+\Omega^2\mp4\Delta\sqrt{\Delta^2+2\Omega^2}}{((32\Delta^2+\Omega^2)(\Delta^2+2\Omega^2)\mp\Delta(32\Delta^2+15\Omega^2)\sqrt{\Delta^2+2\Omega^2})^{1/2}}.   \label{a2}
\end{eqnarray}
\end{widetext}
The corresponding eigenvalues $E_{D_2}=0$ (independent of $\Delta$), and $E_{Dr_\pm}\approx(3\Delta\mp\sqrt{\Delta^2+2\Omega^2})/2$. In Fig. \ref{crossings}(b) we show the dependence of the eigenvalues on detuning $\Delta$ for a strong-interaction case. The eigenvalues solved by numerically diagonalizing $H_{2}$ are represented by the solid lines, and the analytical expressions $E_{D_2}$ and $E_{D_\pm}$ are plotted by the black and magenta dashed lines. 
With a change of detuning, the dark eigenstate $\left\vert D_2\right\rangle$ (black dashed line) is found to coincide with three different numerical eigenstates (green, blue, red solid lines), and especially, at  
\begin{equation}
\Delta=\Delta_{\pm}= \frac{\Omega^2\pm\Omega\sqrt{\Omega^2+4U^2}}{4U},
\label{DeltaC}
\end{equation}
$\left\vert D_2\right\rangle$ becomes degenerate with nearby Rydberg-dressed states $\left\vert Dr_{\pm}\right\rangle$, accompanied by two ACs between the numerical eigenvalues. A partially enlarged view of ACs (shaded area of Fig. \ref{crossings}(b)) is visibly presented in Fig. \ref{crossings}(c). The slight shift between $\left\vert Dr_{\pm}\right\rangle$ and numerical eigenstates (colored solid lines) originates from the condition that $U\gg\Omega$ is not severely met in the calculations.

In the limit of ultra-strong interaction $U\gg\Omega$, Eq. (\ref{DeltaC}) reduces to
\begin{equation}
\Delta=\Delta_{\pm}\approx \pm\frac{\Omega}{2},
\label{ad}
\end{equation}
which is also found by Refs. \cite{Gaul16,Garttner14}, leading to a degeneracy,  $E_{Dr_\pm}=E_{D_2}\approx 0$, there.
Substituting (\ref{ad}) into (\ref{a1}) and (\ref{a2}) straightforwardly comes to saturated excitations $|a_{1\pm}|^2\to 2/3$ and $|a_{2\pm}|^2\to 1/3$. At this case the Rydberg-dressed eigenstates have clear expressions like $\left\vert Dr_{\pm}\right\rangle\approx (\sqrt{2}\left\vert mr\right\rangle_{+}\mp\left\vert mm\right\rangle)/\sqrt{3}$.
On the contrary, when $U$ and $ \Omega$ are comparable, ACs will locate asymmetrically to the center of $\Delta=0$ due to $\Delta_{+}>|\Delta_{-}|$.

\begin{figure}
\includegraphics[width=3.5in,height=2.0in]{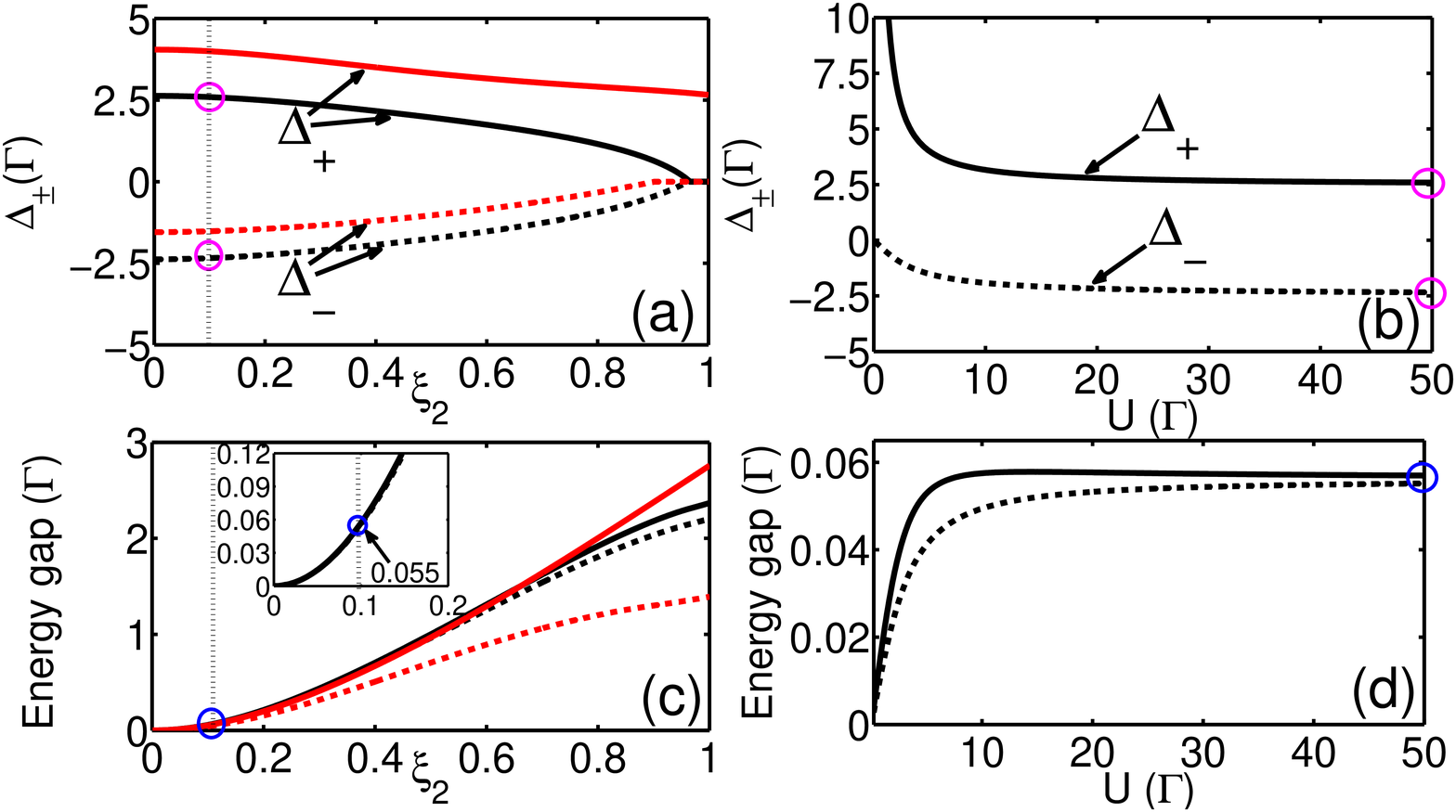}
\caption{(color online). (a) The locations of $\Delta_{\pm}$ versus $\xi_{2}$ for $U=5.0$ (red solid and dash-dotted lines) and $U=50.0$ (black solid and dash-dotted lines). (b) The locations of $\Delta_{\pm}$ versus $U$ for $\xi_{2}=0.1$. Same line-types are used in (c) and (d) with respect to (a) and (b) for the values of energy gaps at ACs. The inset of (c) shows a partially enlarged view for $\xi_{2}\in(0,0.2)$. The case of $U=50$, $\xi_{2}=0.1$ (same as Fig. \ref{crossings}(c)) is pointed out by circles, indicating the energy gap there is 0.055. Other parameters: $\Omega=5.0$, $\Gamma$ is the frequency unit.}
\label{fig2}
\end{figure}

Atoms initially prepared in $\left\vert gg\right\rangle$, as the detuning $\Delta$ adiabatically changes, will experience an excitation facilitation between the degenerate energy levels at $\Delta=\Delta_{\pm}$,  finally reaching the Rydberg dressed states $\left\vert Dr_{\pm}\right\rangle$. 
This excitation process strongly depends on the properties of ACs. Intuitively, it is more facilitated with a larger energy gap when the detuning is scanned from a large negative value across the ACs.
In Fig.~\ref{fig2}, we study the locations and energy gaps of ACs as functions of $\xi_{2}$ and $U$ by solving the secular equation of $H_{2}$. Especially, the values of energy gap are obtained from numerically searching for the minimum difference between the corresponding eigenvalues, well agreeing with our analytical predictions. The points labeled by cycles correspond to the case of Fig. \ref{crossings} where $U=50$, $\xi_2=0.1$ and $\Delta_{\pm}\approx \pm 2.5$, and note that the energy gap for that case is about 0.055 (inset of Fig. \ref{fig2}(c)). In general when the interaction $U$ is strong, two ACs locate symmetrically at $\Delta\approx\pm\Omega/2=\pm2.5$ with equivalent energy gaps. Decreasing $U$ can cause an asymmetrical distribution of $\Delta_{\pm}$ but the relation $\Delta_{+}\geq|\Delta_{-}|$ persists, resulting in a larger energy gap at positive AC. On the other hand, with the increase of $\xi_{2}$ (i.e. the probe laser) two ACs would draw close to one another with enhanced energy gaps. 

\section{Enhanced Excitation based on avoided crossings} 

In the following, we explore the dynamics of a two-atom ensemble, concentrating on the single-excitation collective state. Numerical simulations rely on the master equation:
\begin{equation}
\dot\rho=-i[H_{2},\rho]+ L^{(1)}[\rho]+L^{(2)}[\rho],
\label{mastereq}
\end{equation}
with the Lindblad superoperators given by
\begin{equation}
L^{(i)}=\Gamma\frac{2\sigma_{gm}^{(i)}\rho\sigma_{mg}^{(i)}-\{\sigma_{mm}^{(i)},\rho\}}{2} +\gamma\frac{2\sigma_{gr}^{(i)}\rho\sigma_{rg}^{(i)}-\{\sigma_{rr}^{(i)},\rho\}}{2},
\end{equation}
describing the dissipative process.
Two atoms prepared in $\left\vert gg\right\rangle$($=\left\vert D_{2}\right\rangle$) will evolve into a steady state if the evolution time $t\gg\gamma^{-1}$, $\Gamma^{-1}$. We use observable $F_{r}$ to represent the steady probability of single excitation, that is, exciting one atom into $\left\vert r\right\rangle$ and leaving the other in $\left\vert g\right\rangle$ or $\left\vert m\right\rangle$. Hence $F_{r}= \rho_{gr_+,gr_+}+\rho_{mr_+,mr_+}$. $\rho_{jj}$ is the diagonal element of density matrix $\rho$, which characterizes the population on each state $\left\vert j\right\rangle$ ($\left\vert j\right\rangle\in\Phi$).

\begin{figure}
\includegraphics[width=3.4in,height=2.0in]{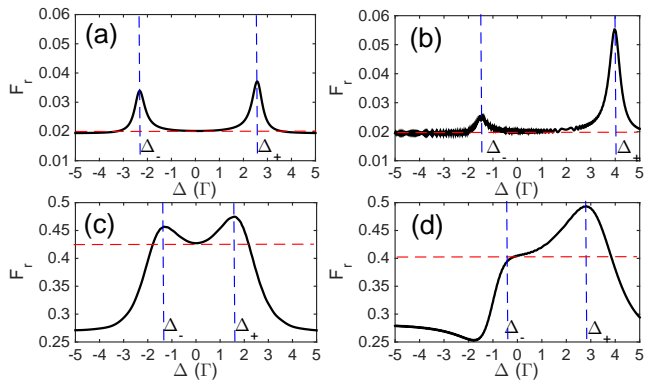}
\caption{(color online). The steady population $F_{r}$ versus detuning $\Delta$. (a) $\xi_{2}=0.1$, $U=50$; (b) $\xi_{2}=0.1$, $U=5.0$; (c) $\xi_{2}=0.5$, $U=50$; (d) $\xi_{2}=0.5$, $U=5.0$. Other parameters are same as used in Fig. \ref{fig2}.}
\label{fr}
\end{figure}

In Fig. \ref{fr}, by solving Eq. (\ref{mastereq}) we obtain the dependence of $F_{r}$ on $\Delta$ under the cases of (a) $\xi_{2}=0.1$, $U=50$, (b) $\xi_{2}=0.1$, $U=5.0$, (c) $\xi_{2}=0.5$, $U=50$, and (d) $\xi_{2}=0.5$, $U=5.0$. When $\xi_{2}$ for the probe laser is small and interaction $U$ is strong, it exhibits an extreme sensitivity to the detuning, having similar Autler-Townes (AT) peaks located at $\Delta_{\pm}\approx\pm\Omega/2$ \cite{DeSalvo16}. As $U$ decreases, the fact that $\Delta_{+}>|\Delta_{-}|$ causes an asymmetrical distribution of two peaks, with the peak values $F_{r}(\Delta_{+})\gg F_{r}(\Delta_{-})$ due to a larger energy gap at positive AC. 
In contrast, if $\xi_{2}$ increases, the two-peak structure is broaden at $\Delta_{\pm}$ and become indiscernible, mainly caused by the resonant excitation for a stronger probe laser. At the same time, the values of $F_{r}$ are found to be increased by one order of magnitude, well coinciding with the variation of energy gap shown in Fig. \ref{fig2}(c). 

Owing to the fact that $F_{r}(\Delta)$ increases significantly with $\xi_{2}$ it is unsuited to characterize the enhancement effect by the exact excitation probability, we have to introduce another relative enhancement factor 
\begin{equation}
EF_{\pm} = F_r(\Delta_{\pm})/F_r(0),
\label{MEF}
\end{equation}
which represents the relative strength of collective excitation at $\Delta=\Delta_{\pm}$ to $\Delta=0$ (resonance). Table \ref{TMEF} summarizes the values of $EF_{\pm}$ for Fig. \ref{fr}. In the first line, all $EF_{\pm}>1$, confirming the fact that applying a constant detuning to the ACs would enhance the excitation; however, in the second line for the intensity-increased probe laser, $EF_{\pm}$ show a trend toward 1.0 (no enhancement), since the resonant excitation is also strongly enhanced at the same time.

\begin{table}[tbp]
\caption{A calculation for relative enhancement factor $EF_{\pm}$ in Fig. \ref{fr}(a)-(d).}%
\begin{tabular}{p{2em}|p{4em}p{4em}|p{2em}|p{4em}p{4em}}
\hline\hline
  & $EF_{+}$ & $EF_{-}$ &  & $EF_{+}$ & $EF_{-}$ \\[6pt]
\hline 
 (a) & 1.83 & 1.67 & (b) & 2.77 & 1.27 \\ 
 (c) & 1.11& 1.07 & (d) & 1.22 & 1.0  \\ 
\hline\hline
\end{tabular}%
\label{TMEF}
\end{table}

\begin{figure}
\includegraphics[width=3.45in,height=1.67in]{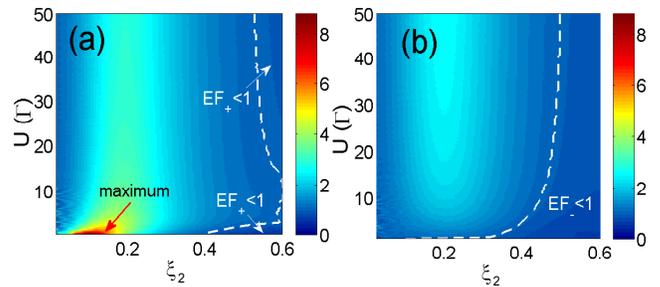}
\caption{(color online). The variation of relative enhancement factor (a) $EF_{+}$ and (b) $EF_{-}$ in the space of $(\xi_{2},U)$. The global maximum of $EF_{+}$($>$8.0) is marked by a red arrow in (a). White dashed lines for $EF_{\pm}=1.0$, label the boundary from suppressed to enhanced excitation. }
\label{MEFfig}
\end{figure}

A thorough value-variation of $EF_{\pm}$ with respect to $U$ and $\xi_{2}$ is displayed in Fig. \ref{MEFfig}. For a larger $U$, it tends to have $EF_{+}\approx EF_{-}$, due to the symmetrical AT peaks. On the other hand, as $\xi_{2}$ increases we clearly see both of $EF_{\pm}$ decrease significantly and even become smaller than unity (no enhancement), as indicated in Fig. \ref{fr}(c-d), irrespective of the values of $U$. The boundary from suppressed to enhanced excitation at $EF_{\pm}=1.0$ is shown by white dashed lines. The global maximum, pointed by a red arrow, appears at the positive-detuning side when both $U$ and $\xi_{2}$ are small. Therefore, in order to observe a visible enhancement in the system of three-level atoms, the key is having an EIT condition ($\omega_{N}\ll\Omega$) combined with a proper detuning ($\Delta=\Delta_{+}$) to the intermediate state $\left\vert m\right\rangle$.

Based on the enhanced excitations at ACs, in the following, we will reveal an anomalous facilitation to the atom-light coupling strength, letting it beyond the usual $\sqrt{N}$ limit as in a strong-blockaded ensemble. For simplicity, we just focus on the case of $\Delta=\Delta_+$ because the enhancement effect is more prominent there.

\section{Beyond $\sqrt{N}$ enhancement} 

In strong-blockaded regime where the suppression of multiple excitations arises due to the strong vdWs interaction, $U\gg\Omega,\omega_{N}$, a $\sqrt{N}$ enhancement to the effective Rabi frequency, that is $\sqrt{N}\Omega_{eff}=(\sqrt{N}\omega_{N})\Omega/2\Delta$ can be obtained. 
In this spirit one can describe the ensemble as a ``superatom", permitting a collective excitation between ground state and entangled states \cite{Honer11}. 
For two such atoms, the enhancement is $\sqrt{2}$. 
Previous works have shown that the limit $\sqrt{N}$ can be overcome by reducing $U$ via enlarging interatomic distance \cite{Beguin13} or a proper electric-field tuning \cite{Qian16}. In this work we find such $\sqrt{N}$ enhancement limit to $\omega_{N}$ of the probe laser could easily be exceeded at $\Delta=\Delta_{+}$ through the dramatically enhanced excitation between the ground and the Rydberg dressed states.

\begin{figure}
\includegraphics[width=3.4in,height=4.7in]{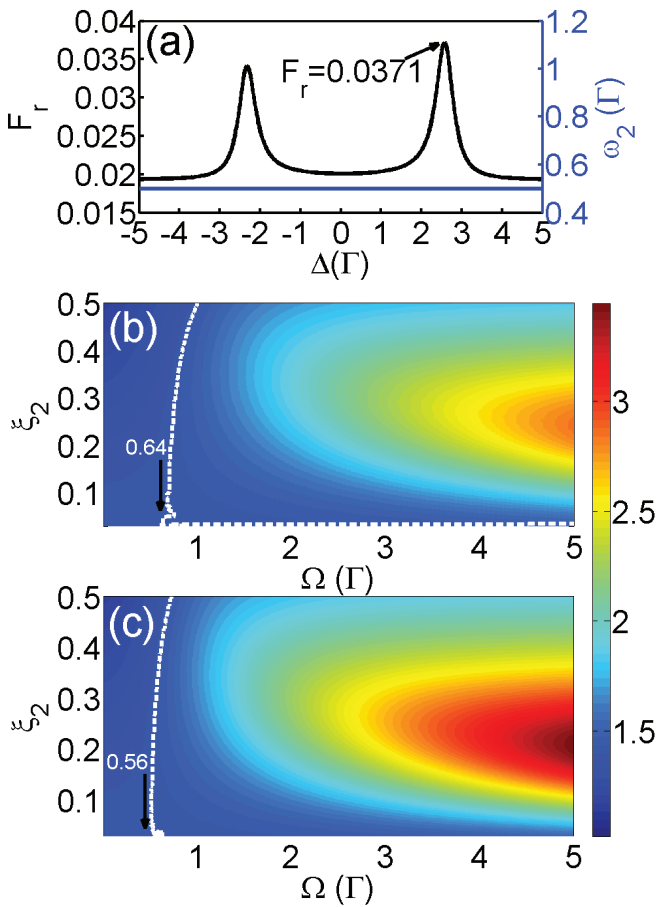}
\caption{(color online). (a) The steady population $F_{r}$ as a function of $\Delta$, which is the same curve as in Fig. \ref{fr}(a). The required value $\omega_{2}$ is shown by a blue solid line. (b) The dependence of $\alpha_{2}$ on $\Omega$ and $\xi_{2}$ in the strong-interaction case $U=10\Omega$. The white dashed line denotes the boundary where $\alpha_{2}=\sqrt{2}$. (c) is the same as (b) except for the weak-interaction case $U=\Omega$. }
\label{EF}
\end{figure}

First, let us introduce a new parameter $\alpha_{N}$, replacing $\sqrt{N}$, to measure the enhancement of the atom-light coupling strength. It is a ratio of the single-atom Rabi frequency to the collective one,
\begin{equation}
\alpha_{N}=\omega_1/\omega_{N},
\label{alpha}
\end{equation}
where $\omega_1=\Omega/\sqrt{f_r^{-1}-1}$ from Eq. (\ref{frvalue}) and $\omega_{N}$ is for the probe laser in a $N$-atom ensemble under the condition $F_{r}=f_{r}$.
The anomalous enhancement can be understood with the help of Fig. \ref{EF}(a), where we re-plot the curve in Fig. \ref{fr}(a) (black solid line) for the case $N=2$ and find the maxima $F_{r}=0.0371$ at $\Delta\approx +2.5$ where the required value of $\omega_{2}$ is about 0.5 (blue solid line). However, in a single-atom frame, $\omega_1=\Omega/\sqrt{f_r^{-1}-1}=0.9814$ and $f_{r}=0.0371$ if same excitation is acquired, giving rise to the enhancement ratio $\alpha_{2}=\omega_{1}/\omega_{2}=1.963>\sqrt{2}$ (the expected value). That means the collective excitation between two interacting atoms can induce an anomalous enhancement to the atom-light coupling strength, making it beyond the value for the ``superatom" \cite{Stanojevic09}.

In the limit $\xi_{2}\to 0$ and $U\gg\Omega$, we find
\begin{equation}
\alpha_{2}\approx\sqrt{2}(1+\xi_{2}^2(1+f(\Omega,\Delta))-(\frac{\Omega}{U}\times g(\Omega,\Delta))),
\label{alpha2}
\end{equation}
which can reduce to $\sqrt{2}$, agreeing with the $\sqrt{N}$ prediction \cite{Ebert14}. In deriving Eq. (\ref{alpha2}) only the lowest order with respect to $\xi_{2}$ and $\Omega/U$ are retained. See Appendix A for detailed expressions of $f(\Omega,\Delta)$ and $g(\Omega,\Delta)$.

In Fig. \ref{EF}, we study the enhancement ratio $\alpha_2$ by varying $\Omega$ and $\xi_{2}$ with (b) $U=10\Omega$ and (c) $U=\Omega$. In  the strong-interaction case (b), white dashed lines denote the boundary of $\alpha_{2}=\sqrt{2}$.
It exhibits that for a strong dissipative case ($\Omega/\Gamma$ is small) $,\alpha_2$ could be smaller than $\sqrt{2}$, identifying a regime where the enhanced excitation does not exist. 
Compared to G\"{a}rttner's work \cite{Garttner14} where the critical condition of enhancement is $\Omega/\Gamma>1.27$, we stress that with an appropriate detuning to the intermediate state, one can get a big breakthrough to that value. The condition becomes $\Omega/\Gamma>0.64$, that significantly relaxes the requirement of the parameter for the excitation enhancement. The fact that $\alpha_{2}\approx \sqrt{2}$ at $\xi_{2}\to 0$ whatever $\Omega$ is, coincides with the analytical expression of Eq. (\ref{alpha2}). 
For comparison, we also display the results for a weak-interaction case in (c), where the limitation for enhancement is further extended to be $\Omega/\Gamma>0.56$, and meanwhile $\alpha_{2}$ is found to be persistently larger than $\sqrt{2}$ even $\xi_{2}\to 0$. That is because the strong-blockade condition is no longer met for the reduced Rydberg-Rydberg interactions.


\section{Extension to a many-atom case} 

\begin{figure}
\includegraphics[width=3.5in,height=2.0in]{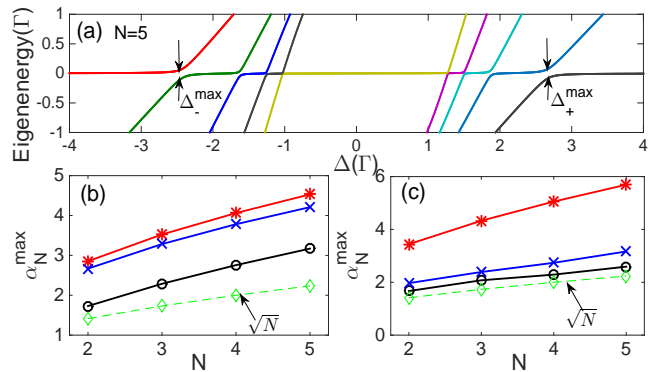}
\caption{(color online). (a) The eigenenergy as a function of $\Delta$ for the case $N=5$, $U=10\Omega$, and $\Delta_{\pm}^{\max}\approx\pm\Omega/2$. Maximal enhancement $\alpha_{N}^{\max}$ for different atomic numbers, $N$=2, 3, 4, 5, are shown for strong-interaction case ($U=10\Omega$) in (b) and weak-interaction case ($U=\Omega$) in (c). The line styles correspond to cases of different detunings, that is, $\Delta$ = $\Delta_{+}^{\max}$ (red line with stars), $\Delta_{-}^{\max}$ (blue line with crosses), and 0 (black line with circles). For comparison, the function $\sqrt{N}$ (green dashed line with diamonds) is displayed in same frame.}
\label{alphaN}
\end{figure}

An intuitive extension is considering an ensemble of $N$ atoms and seeing whether the effect persists as $N$ increases. The inclusion of $N$ atoms gives rise to the emergence of $N-1$ pairs of ACs that respectively locate on the positive and negative sides of detunings, see Fig. \ref{alphaN}(a). 
In the calculations, we directly trace to the pair that possesses the maximal energy gap located at $\Delta_{\pm}^{\max}\approx\pm\Omega/2$ (strong interaction), where a maximal Rydberg excitation probability is expected. For the sake of distinction, the enhancement factor there is denoted by $\alpha_{N}^{\max}$. As in the weak-interaction case, we search for the values of $\Delta_{\pm}^{\max}$ in a numerical way.

Figure \ref{alphaN}(b) and (c) display the dependence of $\alpha_{N}^{\max}$ on the atomic number $N$ for different detunings. 
Generally speaking, $\alpha_{N}^{\max}$ grows with $N$, and at the same time it meets the condition of $\alpha_{N}^{\max}(\Delta_{+}^{\max})>\alpha_{N}^{\max}(\Delta_{-}^{\max})>\alpha_{N}(0)>\sqrt{N}$, irrespective of $N$ and $U$. Changing the interaction $U$ can only affect the relative strengths among them. For example, when (b) $U=10\Omega$, $\alpha_{N}^{\max}(\Delta_{-}^{\max})$ is close to $\alpha_{N}^{\max}(\Delta_{+}^{\max})$ because in strong-interaction case the pair of ACs of maximal gaps will symmetrically distribute at positive and negative sides of detuning. Nevertheless, when (c) $U=\Omega$, $\alpha_{N}^{\max}(\Delta_{-}^{\max})$ tends to be closer to $\alpha_{N}(0)$ since the asymmetry of ACs brings on an ignorable excitation at $\Delta_{-}^{\max}$, e.g. see Fig. \ref{fr}(b). 
In addition, we show that the rate of increase of $\alpha_{N}^{\max}(\Delta_{+})$ grows with $N$, especially in the weak-interaction case due to the significant suppression of interaction at $\Delta=\Delta_{-}^{\max}$. As a result, $\alpha_{5}^{\max}(\Delta_{+})$ reaches as high as $5.7$, far beyond $\sqrt{5}$.

On the flip side of the coin, the enhancement of atom-light coupling also means a reduction of the laser intensity for realizing a same excitation probability, which may be of particular interest for experimentalists. For example, for single-atom case to achieve the excitation $f_{r}= 0.4$ the Rabi frequency of the probe laser should be $\omega_1=4.2$. 
However, for a five-atom ensemble with a weak interaction, $U=\Omega=5$, we obtain the enhancement ratio $\alpha_{5}^{\max}=5.7$ when the detuning $\Delta=4.1$, that means the Rabi frequency $\omega_5=\omega_1/\alpha_5^{\max}=0.73$ for a same excitation probability $F_{r}=0.4$. Therefore the enhancement effect can provide an effective solution to save the intensity of the probe laser in dressing ground-state atoms to the Rydberg levels, which may serve as one-step to the generation of sufficiently strong interactions energy of Rydberg-dressed atom ensemble.

\section{Feasibility and Conclusion}

\begin{table}[tbp]
\caption{A comparison of the realistic parameters used in $^{87}$Rb and $^{84}$Sr atoms for realizing the reduction of the probe laser Rabi frequency in a real implementation. }%
\begin{tabular}{p{7em}|p{9em}|p{9em}}
\hline\hline
  Parameters  & $^{87}$Rb & $^{84}$Sr \\[6pt]
\hline 
 energy levels & $\left\vert g\right\rangle=\left\vert5S_{1/2}\right\rangle$  & $\left\vert g\right\rangle=\left\vert5S_{0}\right\rangle$ \\
  & $\left\vert m\right\rangle=\left\vert5P_{3/2}\right\rangle$ & $\left\vert m\right\rangle=\left\vert5P_{1}\right\rangle$ \\
  & $\left\vert r\right\rangle=\left\vert55S_{1/2}\right\rangle$ \cite{Hofmann13} & $\left\vert r\right\rangle=\left\vert24S_{1}\right\rangle$ \cite{Gaul16} \\
\hline
$ \Gamma/2\pi$ &  6.1 MHz & 76 kHz  \\ 
$ \gamma/2\pi$  &  1.0 kHz &  8.06 kHz   \\
$ U/2\pi$  & 303 MHz  & 24 MHz  \\
$ \Omega/2\pi $ & 30.3 MHz  & 2.4 MHz \\ 
 $\omega_{1}/2\pi$  & 23.1 MHz &  1.83 MHz \\
 $\omega_{5}/2\pi$  & 5.1 MHz    &  165 kHz  \\
 $\alpha_{5}=\omega_1/\omega_5$ & 4.53 & 11.09 \\
  $\Delta/2\pi\approx \Omega/2 $ & 16.7 MHz  & 1.27 MHz  \\
  $F_{r}=f_{r}$  &  0.3675 &  0.3675 \\
  $F_{r}>0.36 $ & $\Delta/2\pi\in(15.4,16.9)$ & $\Delta/2\pi\in(1.257,1.29)$ \\
\hline\hline
\end{tabular}%
\label{feasible}
\end{table}

We check the feasibility of our scheme with realistic parameters of $^{87}$Rb and $^{84}$Sr atoms. The key results and the adopted parameters are summarized in Table \ref{feasible}. For the case of $^{87}$Rb atoms, we find the enhancement ratio $\alpha_5$ can attain 4.53 for the excitation probability $f_{r}=F_{r}=0.3675$, with a detuning $\Delta/2\pi=16.7$MHz to the intermediate state. Specially, we note that in the range of $\Delta\in2\pi\times(15.4,16.9)$MHz, $F_{r}$ maintains a high value ($>0.36$), which indicates that in a realistic experiment, one just has to keep $\Delta\approx\Omega/2$, no need to precisely control the values of detuning.
For comparison, we also study the case of $^{84}$Sr atoms which, confirmed by the experiment \cite{Gaul16}, processes a long-lived middle state $\left\vert m\right\rangle$ ($\Gamma$ is small). For a same excitation probability, the required intensity of probe laser can be reduced by more than 10 times in a five-atom ensemble, and when more atoms are included in the ensemble, a further reduction can be expected.

In conclusion, we perform an extensive study for the anomalous excitation enhancement of Rydberg-dressed atoms under the two-photon resonant EIT condition. In this system we show that there is an approximate dark state with some observable ACs to the Rydberg-dressed states. By adiabatically changing the value of the intermediate-state detuning, the effect of excitation facilitation and its dependence on the atomic interactions as well as the Rabi frequency of probe laser is investigated. We find that this effect will bring on a significant improvement to the atom-probe laser coupling strength, far exceeding the usual $\sqrt{N}$ limit of enhancement predicted in Rydberg-blockade ensemble. In other words, the required intensity of the probe laser for realizing strong interactions between Rydberg-dressed states can be considerably reduced. When more atoms are included, this effect will be further improved. The robustness and feasibility of the scheme are verified by numerical simulations with realistic parameters of $^{87}$Rb and $^{84}$Sr atoms.
Future work will focus on demonstrating the enhanced interactions between Rydberg-dressed states, and the development of new ways for generating Rydberg-dressed atoms with sizable effective interactions as well as long-time coherence based on such effect.

More recently we note that an enhanced excitation induced by Rydberg interaction in a thermal atomic ensemble is experimentally investigated \cite{Kara17}.



\acknowledgements

This work is supported by the NSFC under Grants No. 11474094, No. 11104076, the
Specialized Research Fund for the Doctoral Program of Higher Education No.
20110076120004.

\appendix

\section{}

We give explicit expressions of quantities used in the text, obtained from suitable approximations. In the limit $\xi_{2}\to 0$ and $U\gg\Omega$, we find an analytical expression for the enhancement ratio $\alpha_2$, given by
\begin{equation}
\alpha_{2}\approx\sqrt{2}(1+\xi_{2}^2(1+f(\Omega,\Delta))-(\frac{\Omega}{U}\times g(\Omega,\Delta)))
\end{equation}
where the functions $f(\Omega,\Delta)$ and $g(\Omega,\Delta)$ take forms of

\begin{widetext} 
\begin{eqnarray}
f(\Omega,\Delta)&=&1+\frac{4\Omega^2(800\Delta^4+156\Delta^2\Omega^2+17\Omega^4+50\Delta^2-4\Omega^2)}{(50\Delta^2-4\Omega^2)+(1600\Delta^4+12\Delta^2\Omega^2+25\Omega^4)+4(3200\Delta^6+304\Delta^4\Omega^2+76\Delta^2\Omega^4-\Omega^6)}, \label{f}\\
g(\Omega,\Delta)&=&\frac{2\Delta\Omega(4480\Delta^4+240\Delta^2\Omega^2+43\Omega^4+(280\Delta^2-26\Omega^2))}{(50\Delta^2-4\Omega^2)+(1600\Delta^4+12\Delta^2\Omega^2+25\Omega^4)+4(3200\Delta^6+304\Delta^4\Omega^2+76\Delta^2\Omega^4-\Omega^6)}.   \label{g}
\end{eqnarray}
\end{widetext}

In deducing Eqs. (\ref{f}) and (\ref{g}) we assume $\gamma=0$, all frequencies are in units of $\Gamma$. Obviously, $\alpha_2$ tends to be $\sqrt{2}$ in these limits.

\end{document}